\documentclass[a4paper,11pt]{article}
\usepackage{pos}

\title{The Auger Radio Infill SKALA Extension (ARISE): Science Case and Instrumentation}
\ShortTitle{ARISE: Science Case and Instrumentation}

\manuallySeparateAuthors
\author*[a,b]{Frank G.~Schr\"oder}
\author[c,1]{ for the Pierre Auger Collaboration}

\affiliation[a]{Bartol Research Institute, Department of Physics and Astronomy, University of Delaware, Newark, Delaware, 19716, United States of America}
\affiliation[b]{Institute for Astroparticle Physics (IAP), Karlsruhe Institute of Technology (KIT),\\76021 Karlsruhe, Germany}
\affiliation[c]{Observatorio Pierre Auger, Av. San Martín Norte 304, 5613 Malargüe, Argentina}

\emailAdd{fgs@udel.edu}
\emailAdd{spokerspersons@auger.org}

\note{Full author list at \url{https://www.auger.org/archive/authors_2026_06.html}.}

\abstract{The Auger Radio Infill SKALA Extension (ARISE) at the Pierre Auger Observatory in Argentina was deployed in 2025 and measures cosmic-ray air showers in the energy region of the Galactic-to-extragalactic transition. ARISE is comprised of 18 SKALA-2 antennas featuring two polarization channels each, deployed within $100\,$m of a surface detector station in the enhancement area of the Pierre Auger Observatory. This area of the surface array features a denser spacing of $433\,$m between surface stations, each equipped with underground muon detectors. One of these surface detector stations provides a trigger for simultaneous readout of all ARISE antenna channels. The wide frequency range of ARISE, from $50$ to $350\,$MHz, includes the sub-band of optimum signal-to-noise ratio for air-shower radio emission against the Galactic radio background. In combination with the dense antenna spacing, this enables a relatively low detection threshold, and ARISE aims at demonstrating full detection efficiency for near-vertical air showers above $100\,$ PeV. As an advantage over the current radio detectors at Auger, which are more efficient for inclined air showers, this would enable low systematic uncertainties for physics analysis combining ARISE radio measurements with coincident measurements of the underground muon detectors in the same area. In this presentation, we will provide an overview over the ARISE instrumentation operating at the Pierre Auger Observatory and will outline the science goals.}

\FullConference{11th International Workshop on Acoustic and Radio EeV Neutrino Detection Activities (ARENA2026)\\
8-11 June 2026\\
Karlsruhe, Germany\\}

\begin{document}
\maketitle

\section{Introduction}
The Auger Radio Infill SKALA Extension (ARISE) is a pathfinder array of 18 SKALA-2 antennas~\cite{7297231} deployed in 2025 at the Pierre Auger Observatory in Argentina~\cite{PierreAuger:2015eyc}.
Complementing the Auger Engineering Radio Array (AERA)~\cite{PierreAuger:2025nal} and the Radio Detector (RD) of the AugerPrime upgrade, ARISE aims at full detection efficiency for vertical and near vertical air showers ($\theta \lesssim 40^\circ$) in the energy range of the second knee~\cite{PierreAuger:2023dju}.

\begin{figure}[bh]
    \centering
    \includegraphics[width=0.87\linewidth]{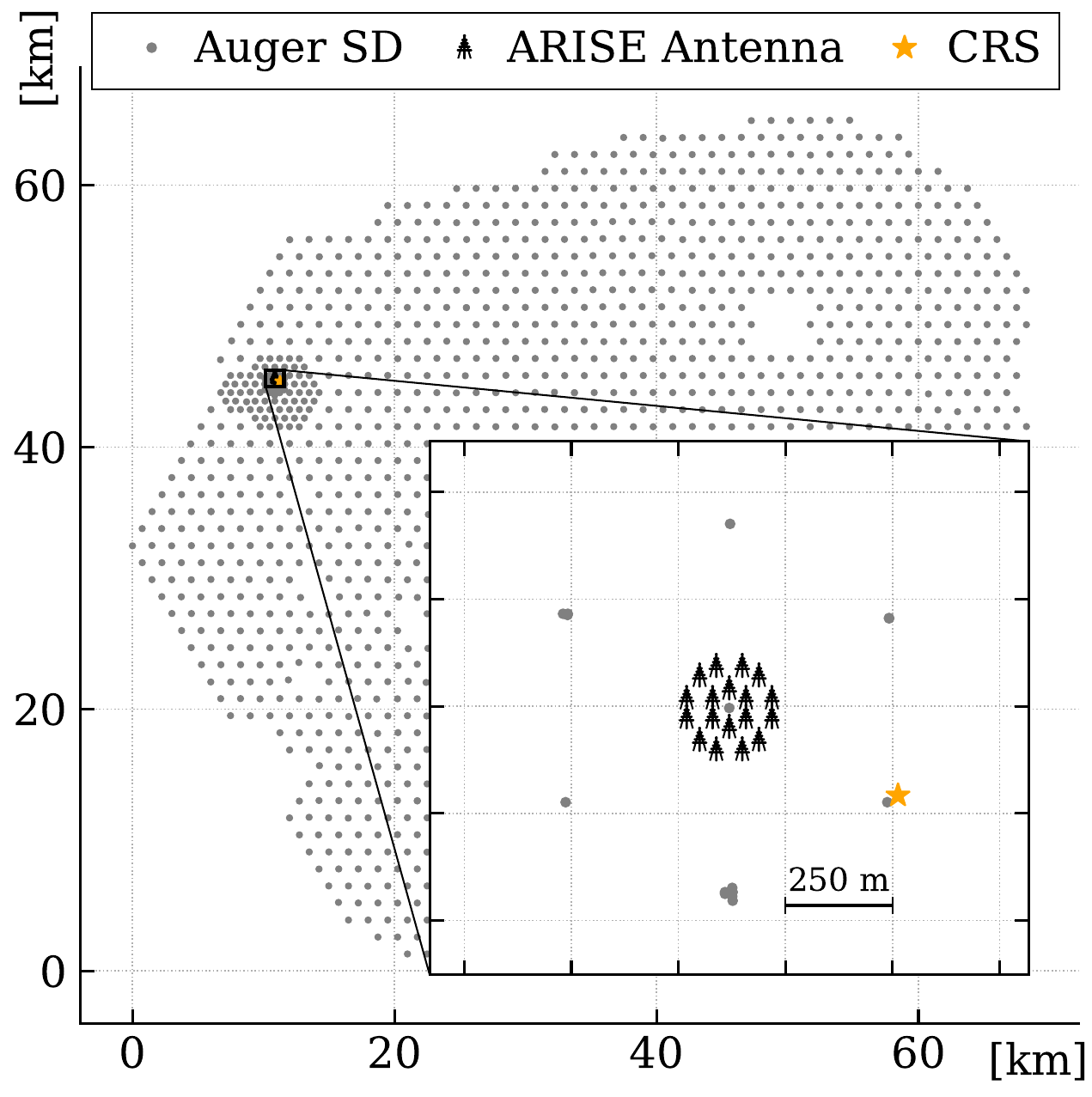} 
    \caption{Map of the surface-detector array of the Pierre Auger Observatory with the locations of the 18 antennas of ARISE (antenna symbols not to scale) and the Central Radio Station (CRS) shown as star.}    
    \label{fig_map}
\end{figure}

The location of ARISE in the densest part of the Auger Muons and Infill for the Ground Array (AMIGA)~\cite{PierreAuger:2021nob}, the SD-433, enables combined radio and muon measurements and technical synergies by housing the central data-acquisition of ARISE in AERA's Central Radio Station (figure~\ref{fig_map}).
The $2\,$km$^2$ SD-433 array of the Auger surface detector is comprises 19 surface detectors (water-Cherenkov detectors with the full AugerPrime upgrades~\cite{Roth:2025hqe}), each equipped with underground muon detectors.
All ARISE antennas are located within approximately $100\,$m radius around one of these surface detectors, which provides the trigger for readout of the radio antennas.

This design therefore automatically provides for coincident measurements of the surface particle, underground muon, and radio signal of air showers, which is an ideal setting for both the technical goal of demonstrating full efficiency for vertical showers and the science goals that rely on combined radio-muon measurements.

This proceeding provides an overview of these science goals and the setup of ARISE.
Separate proceedings report first air-shower measurements of ARISE~\cite{MerxARENA2026} and results from a prototype station of the IceCube-Gen2 surface array in the same area~\cite{VerpoestARENA2026}, which was deployed earlier and provided important expertise for the successful realization of ARISE.

\begin{figure}[t]
    \centering
    \includegraphics[width=0.5\linewidth]{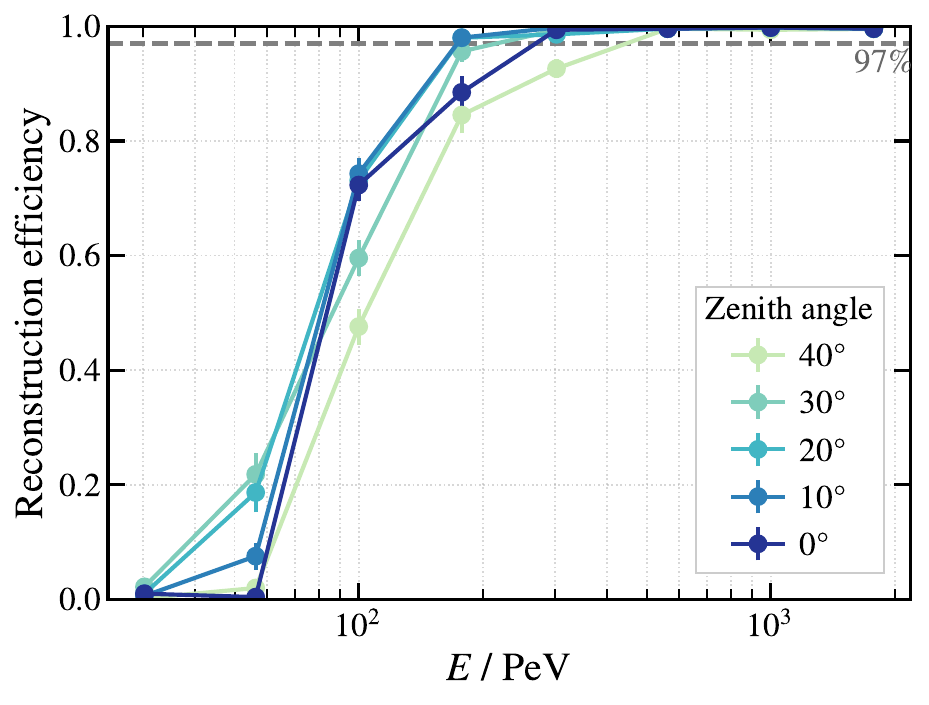}     
    \caption{Simulated efficiency of ARISE for at least five antennas above the detection threshold (from~\cite{Verpoest:2025fyj}).}
    \label{fig_efficiency}
\end{figure}

\section{Goals of ARISE} 

The first goal of ARISE is of a technical nature: a data-driven demonstration of full efficiency for vertical and mildly inclined air showers.
This has yet to be done with current or previous radio arrays for air-shower detection and hence will be a major milestone for the radio technique.
CoREAS simulations with real background measured at the site indicate that full efficiency is expected at a few $10^{17}\,$eV when requiring five antennas above a threshold in signal-to-noise ratio and a successful direction reconstruction (figure \ref{fig_efficiency}, see reference \cite{Verpoest:2025fyj} for details).
Above that energy, every air shower whose axis is contained in the antenna array is expected to be detected, independent of the type of primary particle.

Assuming that neural networks can further reduce the threshold, as recently demonstrated for the same SKALA-2 antennas at IceTop~\cite{IceCube:2025swy}, full efficiency may be achieved already around $10^{17}\,$eV.
In addition to updating simulation studies to the actual ARISE detector configuration (a similar, but ideal layout was used in \cite{Verpoest:2025fyj}), the main goal is to experimentally test the predicted efficiency.
For this purpose, we plan to compare the fraction of triggered air showers for which the radio signal is detected as a function of the energy and zenith angles provided by the SD-433 measurement of the same air showers.

\begin{figure}[t]
    \includegraphics[width=0.48\linewidth]{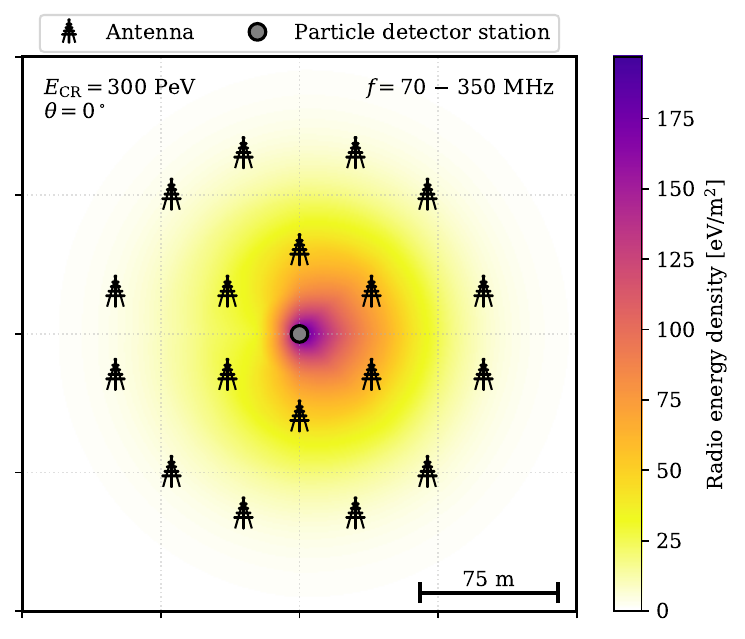} 
    \hfill
    \includegraphics[width=0.48\linewidth]{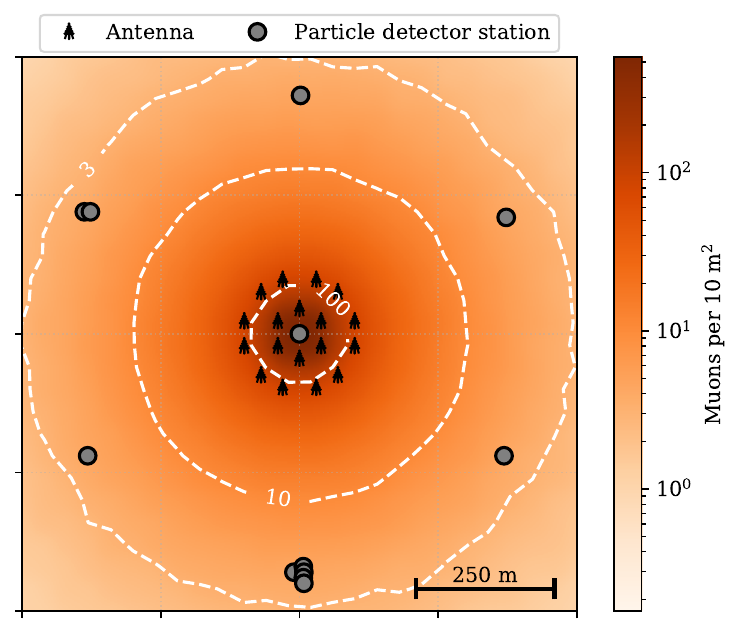}
    \caption{Radio footprint (left) and muon footprint (right) of an air shower on a map of the ARISE array, simulated with CORSIKA. Air showers observed in radio with ARISE are expected to produce a muon signal in most of the surrounding underground muon detectors.}  
    \label{fig_showerFootprint}
\end{figure}

In addition to the SD-433 surface signal, we also plan to combine ARISE radio measurements with the muon measurements of the underground muon detectors in the same area. 
Although the radio footprint of vertical showers is relatively small, the muon footprint is much wider (see figure~\ref{fig_showerFootprint}), so we can expect a measure of the muon density for each ARISE event above the full-efficiency threshold. 
Combining radio and muon measurements is important for maximizing the accuracy of air-shower measurements as $X_\mathrm{max}$ and the size of the muonic shower component provide complementary information on the mass of the primary particle~\cite{Holt:2019fnj,Flaggs:2023exc}.

Full efficiency will then enable lower systematic uncertainties for the following scientific goals:
\begin{itemize}
    \item The coincident measurement with the underground muon detectors will enable studies of hadronic interaction models on an event-by-event level. When reconstructing $X_\mathrm{max}$ by producing many CoREAS simulations and determining which ones match the measured signal best, we will also be able to check if the same simulations are in agreement with the measured muon signal.
    \item The same simulations can also be used to facilitate event-by-event mass separation using the combined radio-muon measurements if, for some measured events, only simulations of certain primary particles reproduce the measurements.
    \item The known mismatch of hadronic interaction models regarding the absolute muon number can potentially be addressed on an average basis as well by statistically calibrating the total range of experimentally observed $X_\mathrm{max}$ values against the range of the muon densities of the same air showers. In addition, this can be used for simultaneous mass composition studies.
    \item As the SKALA-2 antennas are also used at other experiments such as IceCube, ARISE can be used to cross-check the energy scale of Auger's SD-433 and SD-750 array in the energy range of the second knee against IceCube's surface array, IceTop. The method has already been established by a cross-check of the KASCADE-Grande and Tunka energy scales~\cite{Tunka-Rex:2016nto}, indicating that an accuracy of better than $10\,\%$ would have been achievable when using the same antenna type.
\end{itemize}

For all these scientific goals, the best balance of systematic and statistical uncertainties will be determined once the efficiency of ARISE has been experimentally validated. 
It is plausible that at least for some of these goals the total uncertainties can be significantly lowered by increasing the aperture.
Therefore, the current ARISE installation of 18 antennas may eventually serve as a pathfinder for a larger array that covers a greater fraction of the SD-433.
\begin{figure}[t]
    \centering
    \includegraphics[width=0.999\linewidth]{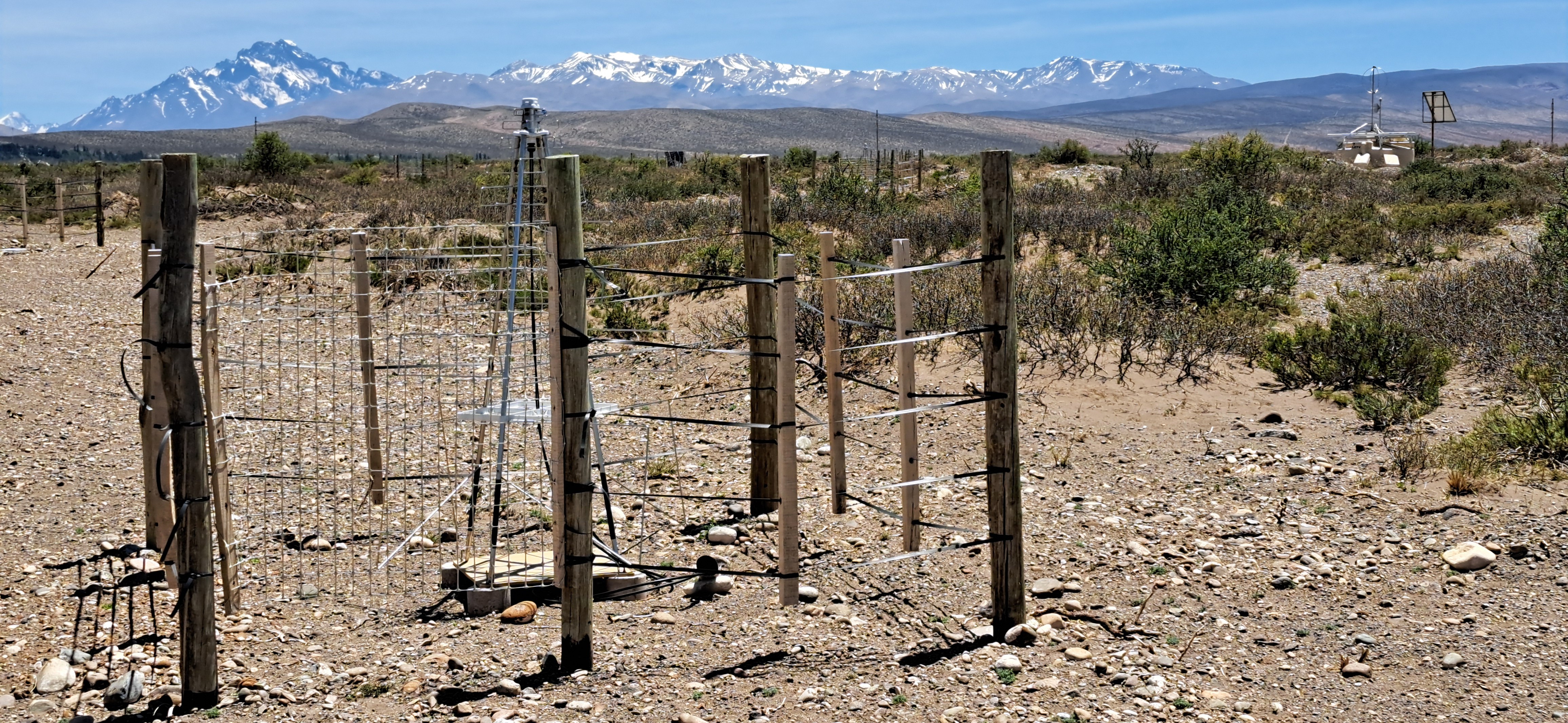} 
    \includegraphics[width=0.999\linewidth]{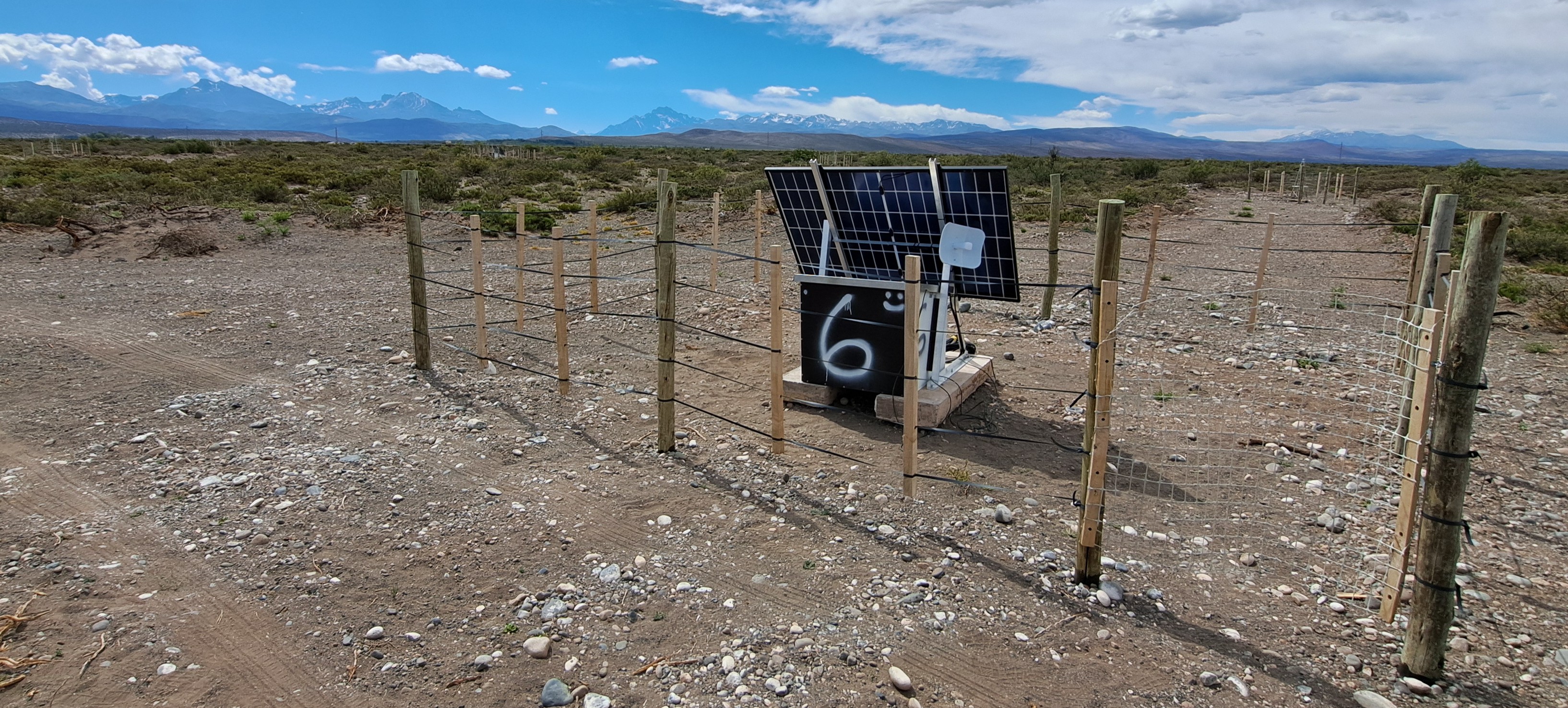}
    \caption{Photos of one of the SKALA-2 antennas of ARISE (top) with SD Lety Jr.~in the background and one of the six station centers of ARISE (bottom) taken in November 2025. Further antennas of ARISE are visible in the background. Non-conductive fences protect from cows and horses living in the area.}
    \label{fig_photos}
\end{figure}

\section{Setup}
ARISE was deployed and started data-taking in 2025. 
The design of the array has been driven by its technical goal to demonstrate full efficiency for vertical showers within the constraint that data-acquisition electronics (DAQ) for six stations of three SKALA-2 antennas each was available.
Based on a simulation study using CoREAS simulations and background of the IceCube-Gen2 prototype station at the Pierre Auger Observatory measured by the same type of DAQ, we therefore concluded that the diameter of ARISE should be about $200\,$m when using a design with hexagonal symmetry \cite{Verpoest:2025fyj}.
For the science goals, ARISE needs to be fully contained in the Auger SD-433 array.
We have thus deployed ARISE symmetrically around one of the surface detector stations, named SD Lety Jr., which provides the trigger simultaneously to all six ARISE stations.

\begin{figure}[b]
    \includegraphics[width=0.99\linewidth]{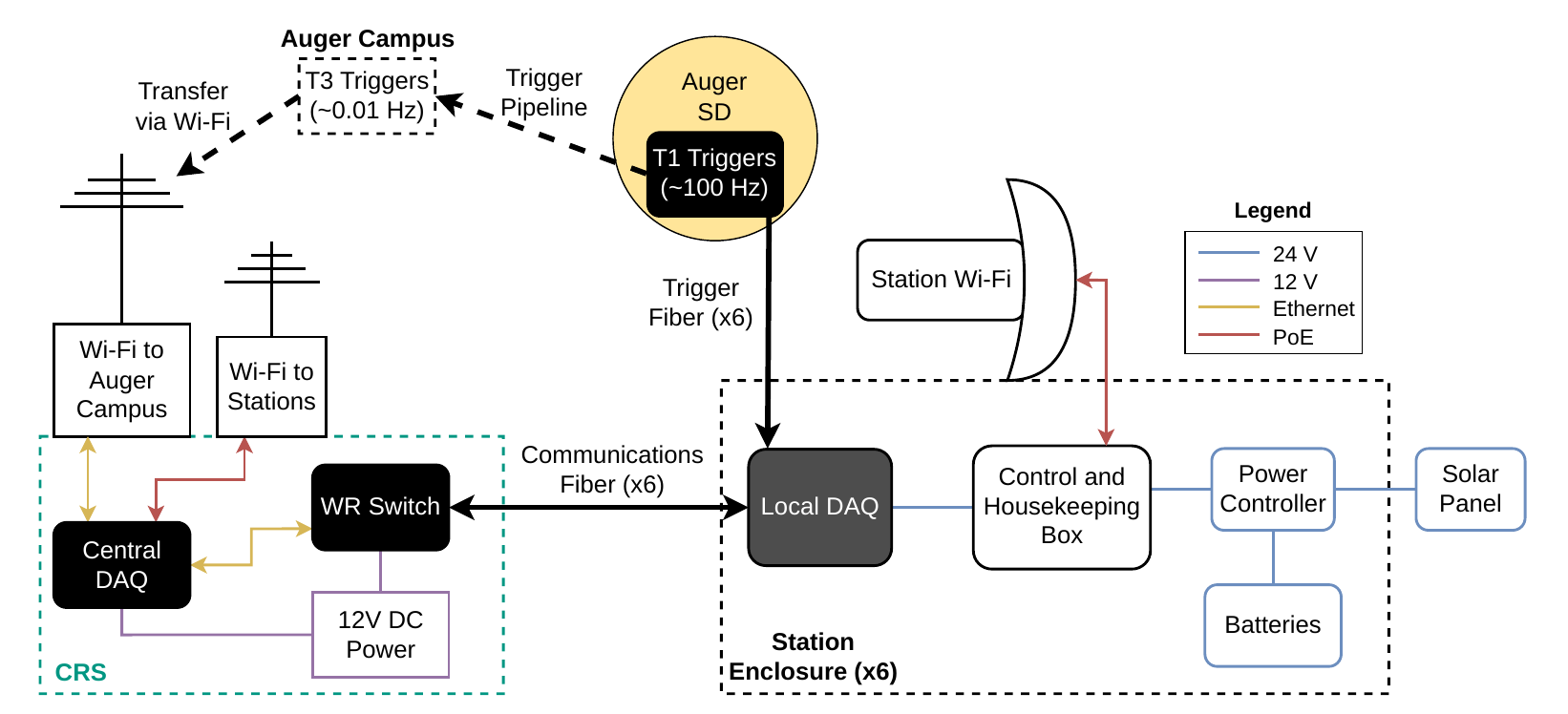} 
    \caption{Schematics of the ARISE data-acquisition as described in the text.}
    
    \label{fig_ARISE_DAQ_schematics}
\end{figure}

Each of the six ARISE stations features the same design of one station center housing the DAQ and three SKALA-2 antennas of two channels each (see figure~\ref{fig_photos}).
Two optical fibers connect each station DAQ, one for the trigger from the Lety Jr.~surface detector station and one from the Central Radio Station (CRS) for timing and data transfer via a White Rabbit switch to the central DAQ, which is a laptop collecting the data from all six ARISE stations. 

The station centers house the local DAQ, which is based on the TAXI design used for the IceTop enhancement~\cite{Shefali:2025icrc,Megha:2025icrc}. 
The main modification with respect to the TAXI v3.2 used in IceCube and for the IceCube-Gen2 prototype station in Auger regards the trigger input. 
Instead of scintillation panels connected to the TAXI, an optical signal generated upon a local trigger (T1) at Lety Jr. initiates the simultaneous readout of all radio channels with a trace length of 2048 samples at $800\,$MSp/s.

\begin{figure}[t]
    \includegraphics[width=0.99\linewidth]{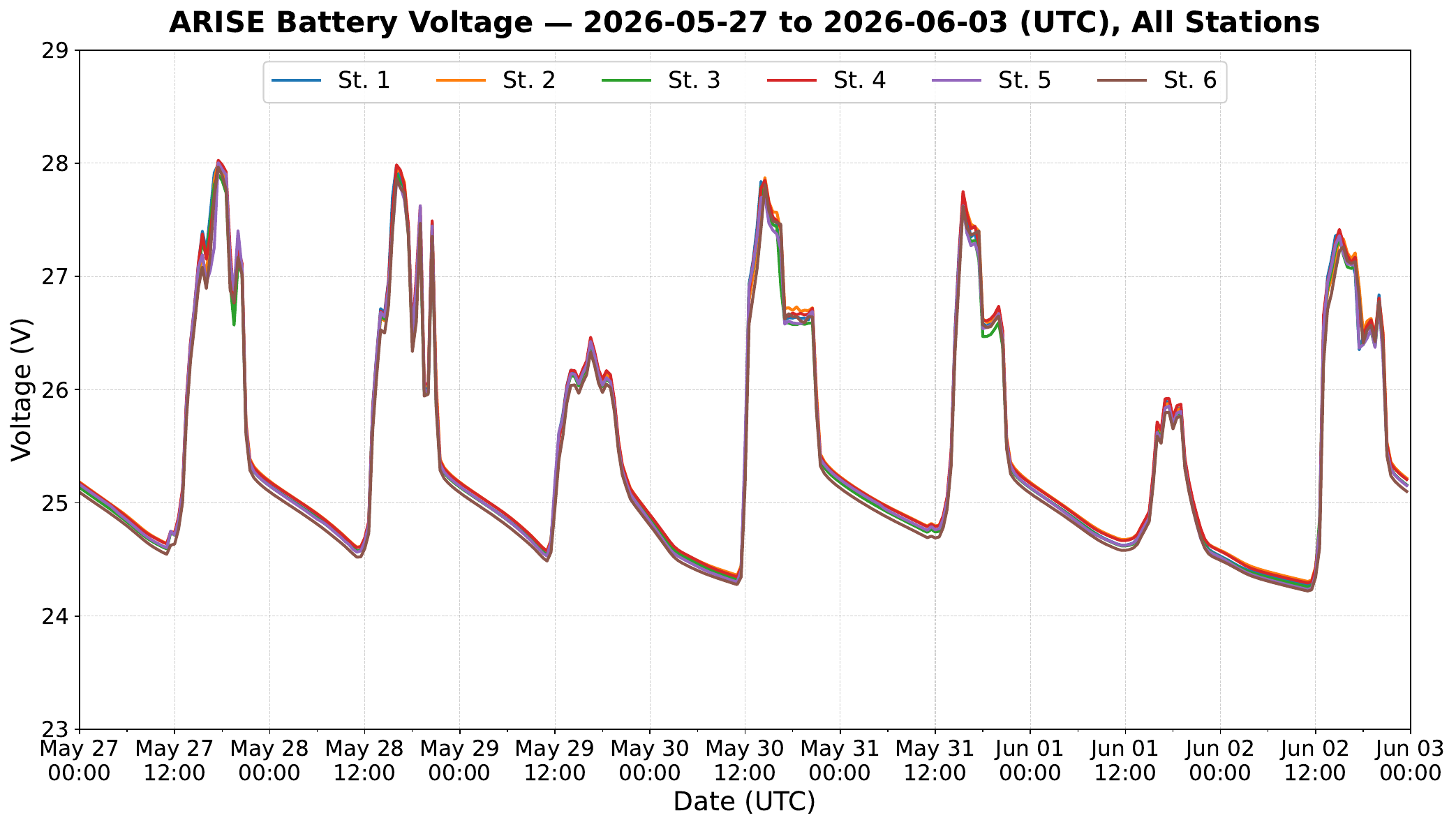} 
    \caption{One week time series of the battery voltage of the six ARISE stations monitored with ARISE CHK: voltages are higher during daytime when batteries are charged by the solar panel, less so during cloudy days. Until the moment of writing, batteries have never reached the shut-down voltage of the charge controller at $23\,$V, which indicates that the battery capacity is sufficient to sustain ARISE operation through several cloudy days even in winter.}    
    \label{fig_batteryMonitoring}
\end{figure}

Solar panels and batteries enable continuous operation, even if cloudy weather persists for several days in winter.
Experience in the commissioning phase showed that the local data taking occasionally stopped, but can be revived by power cycling the station.
Therefore, in March 2026, a Control and House Keeping (CHK) electronics was installed at each station center. 
CHK enables to remotely switch off and on stations through a Wi-Fi connection independent of the optical fibers used for triggering and data transfer.
The installation of CHK has provided for relatively stable data-taking of all ARISE stations since April 2026. 
ARISE CHK also monitors environmental information and the power supply which, together with daily reports of the radio background and trigger rates, provides important monitoring information for the status of ARISE (see figure~\ref{fig_batteryMonitoring}). 

ARISE thereby benefits from the infrastructure put in place for AERA, including a wireless connection from the CRS to the central campus of the Pierre Auger Observatory and access to the internet.
In the CRS, ARISE events are filtered by their time stamps to keep only air-shower events of the Pierre Auger Observatory (defined as an event passing the T3 condition, requiring at least three surface detector stations triggering in time and spatial coincidence). 
This reduces the event rate from $\mathcal{O}(100\,$Hz$)$ to $\mathcal{O}(0.01\,$Hz$)$, enabling the transfer and storage of all merged events.
Currently, they are stored redundantly at the Karlsruhe Institute of Technology and the University of Delaware. 
Figure~\ref{fig_ARISE_DAQ_schematics} provides a schematic of the ARISE setup.

\section{Conclusion}
ARISE has been deployed during the last year, with commissioning data taken in late 2025 and relatively stable data taking of all stations starting in April 2026.
Comprised of 18 SKALA-2 antennas located within the SD-433 array of the Pierre Auger Observatory, its main goal is to demonstrate that a radio array can achieve full efficiency for vertical air showers around $10^{17}\,$eV.
Subsequently, the combined measurements of the ARISE radio antennas and the underground muon detectors can be utilized for a number of science cases in the energy range of the Galactic-to-extragalactic transition.
A first analysis confirms that ARISE is measuring the radio emission from air showers as expected~\cite{MerxARENA2026}.

\bibliographystyle{ICRC}
\bibliography{bibliography}

\section*{Acknowledgments}
We thank Eloy de Lera Acedo and Quentin Gueuning for their support regarding the SKALA-2 antennas.
This project benefited from funding provided by the European Research Council (ERC) and the U.S.~National Science Foundation (NSF). 
Further acknowledgments are included with the full author list at: \url{https://www.auger.org/archive/authors_2026_06.html}

\end{document}